\documentclass[aps,prd,twocolumn,groupedaddress,showpacs,nofootinbib,amssymb]{revtex4}
\usepackage{graphicx,bm,color}
\usepackage{amsmath}
\usepackage{amssymb}
\usepackage{amsfonts}
\usepackage{cases}

\begin{document}

\title{Properties of Bigravity Solutions in a Solvable Class}
\author{Taishi Katsuragawa}
\affiliation{Department of Physics, Nagoya University, Nagoya 464-8602, Japan}

\begin{abstract}
We consider the properties of solutions in the bigravity theory for general models, 
which are parametrized by two parameters $\alpha_{3}$ and $\alpha_{4}$.
Assuming that two metric tensors $g_{\mu \nu}$ and $f_{\mu \nu}$ satisfy the condition
$f_{\mu \nu}=C^{2}g_{\mu \nu}$ where $C$ is a constant,
we investigate the conditions for the parameters so that the solutions with $C\neq1$ could exist.
We also discuss the magnitude and the sign of corresponding cosmological constants.  

For the black hole solution, we consider the black hole entropy to which the massive spin-$2$ field contributes. 
In order to obtain the black hole entropy, we take an approach which uses the Virasoro algebra and the central charge 
corresponding to the surface term in the action.
\end{abstract}


\maketitle

\section{Introduction}
Bigravity is a recently proposed theory, which is nonlinear massive gravity that can be free of ghost 
with the dynamical reference metric 
\cite{Fierz:1939ix,Hinterbichler:2011tt,deRham:2010ik,Hassan:2011hr,Hassan:2011zd,Hassan:2011vm}. 
This gravity model is called bigravity or bi-metric gravity 
because the model contains two symmetric tensor fields $g_{\mu \nu}$ and $f_{\mu \nu}$ and 
a massive spin-$2$ field appears in addition to the massless spin-$2$ field corresponding to the graviton. 
The new degrees of freedom introduced by another metric produce the possibility to solve some problems in cosmology,
that is, dark energy 
\cite{Berg:2012kn,vonStrauss:2011mq,Volkov:2013roa,Akrami:2013ffa,Nojiri:2012re,Nojiri:2012zu,Bamba:2013hza}
and dark matter problems
\cite{Banados:2008fi,Rossi:2009ju}.
The interaction between two metric tensors produces the cosmological constant effectively, 
furthermore, the matter coupled with the metric $f_{\mu \nu}$ produces new gravitational sources as dark matter. 

Besides cosmological applications, bigravity is also interesting as a model of higher spin field theory
\cite{Hinterbichler:2012cn,Boulanger:2000rq,Isham:1971gm}
because the bigravity model describes the massive spin-$2$ field coupled to gravity.
There is actually a spin-$2$ meson as massive spin-$2$ field in QCD, 
and massive spin-$2$ fields are predicted from string theory or higher-dimension theories.

Although it is important to understand the properties of the bigravity, 
it is not easy in general to investigate the solutions because of two reasons.
One reason is that bigravity contains too many parameters. 
These parameters define the form of interactions, and the solutions depend on its combination.  
Another is that we have too many degrees of freedom in two tensor fields.
We have to solve the dynamics of two tensor fields simultaneously, because they are interacting with each other. 

In this work, we consider the general model of bigravity without specifying the parameters,
and study the properties of the solutions.
In order to obtain the solution, we only assume that one metric is proportional to another,
$f_{\mu \nu}=C^{2}g_{\mu \nu}$ with a constant $C$.
We investigate the parameter region which gives non-trivial solutions with $C\neq1$,
and we also discuss the magnitude of cosmological constant and the condition under which cosmological constant vanishes.

Furthermore, we consider the black hole solutions in bigravity and evaluate the black hole entropy.
As mentioned above, bigravity describes the massive spin-$2$ field coupled to gravity.
Therefore, by considering the black holes in bigravity, we can evaluate how the massive spin-$2$ 
field near the horizon affects the black hole entropy
\cite{Brito:2013wya,Volkov:2012wp,Banados:2011hk,Banados:2011np}.
The author has evaluated the black hole entropy by using the holography \cite{Katsuragawa:2013bma},
where we found the solution $f_{\mu \nu}=g_{\mu \nu}$ for a minimal model,
and obtain the entropy which is twice as much as the Bekenstein-Hawking entropy in the Einstein gravity. 

In this work, 
we show that the entropy is given by a sum of two entropies which stem from two Ricci scalars.
For black hole solution, we use the results obtained by the above analysis in general model,
and evaluate the contribution from the massive spin-$2$ field which affects the Bekenstein-Hawking entropy.

\section{Equation of Motion for Bigravity}

The action of bigravity is given by
\begin{align} 
&S_\mathrm{bigravity} \nonumber \\ 
& \quad = M^{2}_{g} \int \, d^{4}x \sqrt{-\mathrm{det}(g)}R(g) \nonumber \\
& \quad + M^{2}_{f} \int \, d^{4}x \sqrt{-\mathrm{det}(f)}R(f)  \nonumber \\
& \quad - 2m^{2}_{0} \, M^{2}_\mathrm{eff} \int \, d^{4}x \sqrt{-\mathrm{det}(g)} 
\sum^{4}_{n=0} \beta_{n}e_{n} \left( \sqrt{g^{-1}f} \right)  \label{the action}
\end{align}
Here, $g$ and $f$ are dynamical variables and rank-two tensor fields which have properties as metrics,
$R(g)$ and $R(f)$ are the Ricci scalars for $g_{\mu \nu}$ and $f_{\mu \nu}$, respectively,
$M_{g}$ and $M_{f}$ are the two Planck mass scales for $g_{\mu \nu}$ and $f_{\mu \nu}$ as well, 
and the scale $M_\mathrm{eff}$ is the effective Planck mass scale defined by
\begin{align}
\frac{1}{M^{2}_\mathrm{eff}} = \frac{1}{M^{2}_{g}} + \frac{1}{M^{2}_{f}}.
\label{Meff}
\end{align}
The quantities $\beta_{n}$s and $m_{0}$ are free parameters, 
and the former defines the form of interactions and the latter expresses the mass of the massive spin-$2$ field.
The matrix $\sqrt{g^{-1}f}$ is defined by the square root of $g^{\mu \rho}f_{\rho \nu}$, that is,
\begin{align}
\left( \sqrt{g^{-1}f} \right)^{\mu}_{\ \rho} \left( \sqrt{g^{-1}f} \right)^{\rho}_{\ \nu} = g^{\mu \rho}f_{\rho \nu}. \label{sqrtfg}
\end{align}
For general matrix $\mathbf{X}$, $e_{n}(\mathbf{X})$s are polynomials of the eigenvalues of  $X$:
\begin{align}
e_{0}(\mathbf{X}) =& 1 , \quad 
e_{1}(\mathbf{X}) = [\mathbf{X}] , \nonumber \\
e_{2}(\mathbf{X}) =& \frac{1}{2} \left( [\mathbf{X}]^{2} - [\mathbf{X}^{2}] \right) , \nonumber \\
e_{3}(\mathbf{X}) =& \frac{1}{6} 
\left( [\mathbf{X}]^{3} - 3[\mathbf{X}][\mathbf{X}^{2}] + 2[\mathbf{X}^{3}] \right) , \nonumber \\
e_{4}(\mathbf{X}) =& \frac{1}{24} 
\left( [\mathbf{X}]^{4} - 6[\mathbf{X}]^{2}[\mathbf{X}^{2}] + 3[\mathbf{X}^{2}]^{2} \right. \nonumber \\
&+ \left.  8[\mathbf{X}][\mathbf{X}^{3}] - 6[\mathbf{X}^{4}]  \right)  \nonumber \\
=& \mathrm{det}(\mathbf{X}) , \nonumber \\ 
e_{k}(\mathbf{X}) =& 0 \quad \mbox{for} \ \ k>4 , \label{e_n}
\end{align}
where the square brackets denote traces of the matrices, that is, $[X]=X^{\mu}_{\mu}$.
For conventional notation, we explicitly denote the determinant of matrix $A$ as $\mathrm{det}(A)$, 
and $\sqrt{A}$ represents matrix which is the square root of $A$.

Now we consider the variation of the action (\ref{the action}) with respect to $g_{\mu \nu}$,
\begin{align}
&\delta_{g} S_\mathrm{bigravity} \nonumber \\
& = M^{2}_{g} \int \, d^{4}x \delta_{g}\left( \sqrt{-\mathrm{det}(g)}R(g) \right) \nonumber \\
& - 2m^{2}_{0} \, M^{2}_\mathrm{eff} \int \, d^{4}x \delta_{g} \left( \sqrt{-\mathrm{det}(g)} 
\sum^{4}_{n=0} \beta_{n}e_{n} \left( \sqrt{g^{-1}f} \right) \right) \label{variation} .
\end{align}
The first term produces the well-known Einstein tensor,
\begin{align}
&\delta_{g}\left( \sqrt{-\mathrm{det}(g)}R(g) \right) \nonumber \\
& \quad = \sqrt{-\mathrm{det}(g)} \left( R_{\mu \nu}(g) - \frac{1}{2}R(g)g_{\mu \nu} \right) \delta g^{\mu \nu} \nonumber \\
& \quad + \mbox{total derivative terms} \label{variation1}.
\end{align}
The second term produces the interactions between $g_{\mu \nu}$ and $f_{\mu \nu}$, 
\begin{align}
&\delta_{g} \left( \sqrt{-\mathrm{det}(g)} \sum^{4}_{n=0} \beta_{n}e_{n} \left( \sqrt{g^{-1}f} \right) \right) \nonumber \\
&=\delta_{g} \left( \sqrt{-\mathrm{det}(g)} \sum^{3}_{n=0} \beta_{n}e_{n} \left( \sqrt{g^{-1}f} \right)  
+ \beta_{4}\sqrt{-\mathrm{det}(f)}\right) \nonumber \\
&=\sqrt{-\mathrm{det}(g)} 
\left( - \frac{1}{4} \sum^{3}_{n=0}(-1)^{n}\beta_{n} \right. \nonumber \\
&\times \left. \left[ g_{\mu \lambda}Y^{\lambda}_{(n) \nu}(\sqrt{g^{-1}f}) 
+ g_{\nu \lambda}Y^{\lambda}_{(n) \mu}(\sqrt{g^{-1}f}) \right] \right) \delta g^{\mu \nu}
\label{variation2} .
\end{align} 
In the second line of (\ref{variation2}), 
we have used the following property,
\begin{align}
\sqrt{\mathrm{det}(-g)}\, \beta_{4}\sqrt{\mathrm{det}(g^{-1}f)}
=\beta_{4}\sqrt{\mathrm{det}(-f)} ,
\end{align} 
and the variation vanishes because it does not depend on $g_{\mu \nu}$.
So, the equation of motion for $g_{\mu \nu}$ is given by
\begin{align}
0 =& R_{\mu \nu}(g) - \frac{1}{2}R(g)g_{\mu \nu} \nonumber \\
&+ \frac{1}{2} \left( \frac{m_{0}M_\mathrm{eff}}{M_{g}} \right)^{2}
\sum^{3}_{n=0}(-1)^{n}\beta_{n} \nonumber \\ 
&\times \left[ g_{\mu \lambda}Y^{\lambda}_{(n) \nu}(\sqrt{g^{-1}f}) 
+ g_{\nu \lambda}Y^{\lambda}_{(n) \mu}(\sqrt{g^{-1}f}) \label{geq} \right] 
\end{align}
Here, for a matrix $\mathbf{X}$, $Y_{n}(\mathbf{X})$s are defined by
\begin{align}
Y^{\lambda}_{(n) \nu}(\mathbf{X})=\sum^{n}_{r=0} (-1)^{r} \left( X^{n-r} \right)^{\lambda}_{\ \nu} e_{r}(\mathbf{X}) ,
\end{align}
or explicitly, 
\begin{align}
Y_{0}(\mathbf{X}) =& \mathbf{1} , \quad 
Y_{1}(\mathbf{X}) = \mathbf{X}- \mathbf{1}[\mathbf{X}] , \nonumber \\
Y_{2}(\mathbf{X}) =& \mathbf{X}^{2} - \mathbf{X}[\mathbf{X}] 
+ \frac{1}{2} \mathbf{1} \left( [\mathbf{X}]^{2} - [\mathbf{X}^{2}] \right) , \nonumber \\
Y_{3}(\mathbf{X}) =& \mathbf{X}^{3} - \mathbf{X}^{2}[\mathbf{X}] 
+ \frac{1}{2} \mathbf{X} \left( [\mathbf{X}]^{2} - [\mathbf{X}^{2}] \right) \nonumber \\
&- \frac{1}{6} \mathbf{1} \left( [\mathbf{X}]^{3} - 3[\mathbf{X}][\mathbf{X}^{2}] + 2[\mathbf{X}^{3}] \right). 
\end{align}
Note that $e_{n}$s are written in terms of the trace of $g^{-1}f$, and it is useful to consider the variation of the trace as follows:
\begin{align}
\delta \mathrm{tr}\left( (\sqrt{g^{-1}f} )^{n} \right) 
= \frac{n}{2} \mathrm{tr} \left( g (\sqrt{g^{-1}}f)^{n} \delta g^{-1} \right) .
\end{align}
Then, we obtain
\begin{align}
&\frac{2}{\sqrt{- \mathrm{det}(g)}} 
\delta_{g} \left( \sqrt{- \mathrm{det}(g)}e_{n}(\sqrt{g^{-1}f}) \right) \nonumber \\
&=\sum^{n}_{r=0}(-1)^{r+1} \mathrm{tr} \left( g (\sqrt{g^{-1}f})^{r} \delta g^{-1} \right) e_{n-r}(\sqrt{g^{-1}f}) ,
\end{align}
and the third line of Eq.(\ref{variation2}) is symmetrized with respect to the indices $\mu$ and $\nu$ .

We need not only the equation of motion for $g$ but that of $f$ because $f$ is dynamical as well as $g$ in bigravity.
In order to obtain the equation of motion for $f_{\mu \nu}$, we can utilize the following symmetry,
\begin{align}
&\sqrt{-\mathrm{det}(g)} \sum^{4}_{n=0} \beta_{n}e_{n} \left( \sqrt{g^{-1}f} \right) \nonumber \\
&=\sqrt{-\mathrm{det}(f)} \sum^{4}_{n=0} \beta_{4-n}e_{n} \left( \sqrt{f^{-1}g} \right) ,
\end{align}
that is,
\begin{align}
g_{\mu \nu} \leftrightarrow f_{\mu \nu} , \quad m_{g} \leftrightarrow m_{f} , \quad \beta_{n} \leftrightarrow \beta_{n-4} .
\end{align}
Applying this symmetry to Eq.(\ref{geq}), we find the equation of motion for $f_{\mu \nu}$ is given by
\begin{align}
0 =& R_{\mu \nu}(f) - \frac{1}{2}R(f)f_{\mu \nu} \nonumber \\
&+ \frac{1}{2} \left( \frac{m_{0}M_\mathrm{eff}}{M_{f}} \right)^{2} 
\sum^{3}_{n=0}(-1)^{n}\beta_{4-n} \nonumber \\
&\times \left[ f_{\mu \lambda}Y^{\lambda}_{(n) \nu}(\sqrt{f^{-1}g}) 
+ f_{\nu \lambda}Y^{\lambda}_{(n) \mu}(\sqrt{f^{-1}g}) \right] . \label{feq}
\end{align}

Now we have obtained the equations of motion for $g$ and $f$ in general model of bigravity, 
and we can investigate the properties of two metrics by solving two equations.
It is difficult, however, to solve the equations and to obtain analytic solution without any assumption
because we have too many degrees of freedom for $g$ and $f$;
naively, bigravity contains twice degrees compared with the general relativity 
since two fields $g$ and $f$ are independent.

From the viewpoint of finding the solutions, we have some lessons from the general relativity.
In the general relativity, we impose some assumptions for the spacetime to find an exact solution.
For example,
we obtain the Schwaraschild solution for static and spherically symmetric spacetime.
In the following section, we assume that two metrics are related to each other, 
and two dynamics for two tensor fields are not independent with each other.

\section{Proportionally-related Solutions and Their Properties}

\subsection{Equations of Motion}

Now, we consider the case where $f_{\mu \nu}=C^{2}g_{\mu \nu}$ and $C$ is a constant
\cite{Hassan:2012wr,Hassan:2012rq}.
This assumption is simple to solve equations of motion
because we have only to determine one tensor field and one constant rather than two tensor fields.
Furthermore, considering the interaction between two metric tensors, 
it might be reasonable to assume that the metrics get identical configuration dynamically,
and two metric tensors may be proportional to each other.

By using the assumption, we obtain 
\begin{align}
&\sqrt{g^{-1}f} = |C|\mathbf{1} , & &[\sqrt{g^{-1}f}] = 4|C| , \nonumber \\
&\sqrt{f^{-1}g} = |C|^{-1}\mathbf{1} , & &[\sqrt{f^{-1}g}] = 4|C|^{-1} 
\end{align}
and corresponding $Y_{n}$s are given by
\begin{align}
&\left\{
\begin{array}{l}
Y_{0}(\sqrt{g^{-1}f}) = \mathbf{1}  , \\
Y_{1}(\sqrt{g^{-1}f}) = -3|C|\mathbf{1} , \\
Y_{2}(\sqrt{g^{-1}f}) = 3C^{2}\mathbf{1} , \\
Y_{3}(\sqrt{g^{-1}f}) = -C^{2}|C|\mathbf{1} \\
\end{array}
\right.  \\
&\left\{
\begin{array}{l}
Y_{0}(\sqrt{f^{-1}g}) = {\bf 1} , \\
Y_{1}(\sqrt{f^{-1}g}) = -3|C|^{-1}{\bf 1}  , \\
Y_{2}(\sqrt{f^{-1}g}) = 3C^{-2}{\bf 1} , \\
Y_{3}(\sqrt{f^{-1}g}) = -C^{-2}|C|^{-1}{\bf 1} .
\end{array}
\right. 
\end{align}
Now, we obtain the two Einstein equations with cosmological constant as follows:
\begin{align}
0 =& R_{\mu \nu}(g) - \frac{1}{2}R(g)g_{\mu \nu} + \Lambda_{g}(C)g_{\mu \nu}  \label{geq1} \\
0 =& R_{\mu \nu}(f) - \frac{1}{2}R(f)f_{\mu \nu} + \Lambda_{f}(C) f_{\mu \nu} \label{feq1} ,
\end{align}
and two cosmological constants are defined as follows:
\begin{align}
\Lambda_{g}(C) =& \left( \frac{m_{0}M_\mathrm{eff}}{M_{g}} \right)^{2} \nonumber \\
& \times \left[ \beta_{0} + 3|C|\beta_{1} + 3C^{2}\beta_{2} + C^{2}|C|\beta_{3}  \right] \\
\Lambda_{f}(C) =& \left( \frac{m_{0}M_\mathrm{eff}}{M_{f}} \right)^{2} \frac{1}{C^{2}|C|} \nonumber \\
& \times \left[ \beta_{1} + 3|C|\beta_{2} + 3C^{2}\beta_{3} + C^{2}|C|\beta_{4} \right] .
\end{align}
Here the dynamics of two metric tensors $g_{\mu \nu}$ and $f_{\mu \nu}$ are separated from each other, 
and the Bianchi identity is automatically satisfied.
This structure of dynamics means that if $f=C^{2}g$, 
the solutions of bigravity is those of the general relativity, 
and we can use the solutions in the general relativity.

Now, we express five $\beta_{n}$s in terms of two free parameters $\alpha_{3}$ and $\alpha_{4}$
\cite{Hassan:2011vm}, as follows:
\begin{align}
&\beta_{0} = 6 - 4 \alpha_{3} + \alpha_{4} , \quad
\beta_{1} = -3 + 3\alpha_{3} - \alpha_{4} \nonumber \\
&\beta_{2} = 1- 2\alpha_{3} + \alpha_{4} , \quad
\beta_{3} = \alpha_{3} - \alpha_{4} , \quad
\beta_{4} = \alpha_{4} .
\end{align}
Also, we can take $C>0$ without loss of generality.
Then we obtain
\begin{align}
\Lambda_{g}(C) 
=& \left( \frac{m_{0}M_\mathrm{eff}}{M_{g}} \right)^{2} \nonumber \\ 
&\times \left[ (6 - 4 \alpha_{3} + \alpha_{4}) + 3C(-3 + 3\alpha_{3} - \alpha_{4}) \right. \nonumber \\
& \left. + 3C^{2}(1- 2\alpha_{3} + \alpha_{4}) + C^{3}(\alpha_{3} - \alpha_{4})  \right]  \nonumber \\
=& \left( \frac{m_{0}M_\mathrm{eff}}{M_{g}} \right)^{2} (C-1) \nonumber \\
&\times \left [ (\alpha_{3}-\alpha_{4})C^2 + (-5\alpha_{3} + 2\alpha_{4} +3)C \right. \nonumber \\
& \left. \quad + (4\alpha_{3} - \alpha_{4} -6) \right ] , \label{lambdag} \\
\Lambda_{f}(C) 
=& \left( \frac{m_{0}M_\mathrm{eff}}{M_{f}} \right)^{2} \frac{1}{C^{3}} \nonumber \\ 
& \left[ (-3 + 3\alpha_{3} - \alpha_{4}) + 3C(1- 2\alpha_{3} + \alpha_{4}) \right. \nonumber \\
& \left. + 3C^{2}(\alpha_{3} - \alpha_{4}) + C^{3}\alpha_{4} \right] \nonumber \\
=& \left( \frac{m_{0}M_\mathrm{eff}}{M_{f}} \right)^{2} \frac{C-1}{C^{3}} \nonumber \\
&\times \left [ \alpha_{4}C^2 + (3\alpha_{3} -2\alpha_{4})C + (-3\alpha_{3} + \alpha_{4} + 3) \right ] . \label{lambdaf}
\end{align}
For the consistency, both of Eqs.(\ref{geq1}) and (\ref{feq1}) should be identical with each other. 
By putting $ f_{\mu \nu} = C^{2} g_{\mu \nu}$, we find $R_{\mu \nu}(f)=R_{\mu \nu}(g)$, 
$R(f)f_{\mu \nu}=R(g)g_{\mu \nu}$. 
Then, we find 
\begin{align}
\Lambda_{g}=C^{2}\Lambda_{f} ,
\end{align}

From the Eqs.(\ref{lambdag}) and (\ref{lambdaf}), we obtain the quartic equation as follows:
\begin{align}
0=&(C-1) \nonumber \\
\times & \left [ M^{2}_{\mathrm{ratio}}(\alpha_{3}-\alpha_{4})C^{3} \right. \nonumber \\
& + \{ -5M^{2}_{\mathrm{ratio}}\alpha_{3} + (2M^{2}_{\mathrm{ratio}} - 1)\alpha_{4} 
+3M^{2}_{\mathrm{ratio}} \}C^{2} \nonumber \\
& + \{ (4M^{2}_{\mathrm{ratio}} -3)\alpha_{3} - (M^{2}_{\mathrm{ratio}} - 2)\alpha_{4} 
- 6M^{2}_{\mathrm{ratio}} \}C \nonumber \\ 
& \left. + (3\alpha_{3} - \alpha_{4} - 3) \right ] \label{Ceq} ,
\end{align}
where we define $M_{\mathrm{ratio}}\equiv M_{f}/M_{g}$. 

Apparently, we can find that general model with arbitrary $\alpha_{3}$ and $\alpha_{4}$ has solution 
where $f_{\mu \nu}=g_{\mu \nu}$, that is $C=1$, and therefore two cosmological constants vanish,
which tells that general model in bigravity has the solution $g_{\mu \nu}=f_{\mu \nu}$ 
which is asymptotically flat solution in the general relativity.
Now, we concentrate on the cubic part in Eq.(\ref{Ceq}) 
and classify two parameters $\alpha_{3}$ and $\alpha_{4}$ when $C\neq1$.
If there is no solution which satisfies $C>0$ and $C\neq1$, 
we do not have non-trivial solution in bigravity.

\subsection{Classification of Solutions}\label{classification}
As a first step, we classify the parameter region by the existence of the solutions.
In the following, we consider the case $M_{f}=M_{g}$, that is, $M_{\mathrm{ratio}}=1$ for simplicity, 
and define a function $F_{3}(x)$ as follows:
\begin{align}
F_{3}(x) 
\equiv & (\alpha_{3}-\alpha_{4})x^{3} - (5\alpha_{3} - \alpha_{4} - 3) x^{2} \nonumber \\
& + (\alpha_{3} + \alpha_{4} - 6)x + (3\alpha_{3} - \alpha_{4} - 3) . \label{cubic}
\end{align}
We now solve the equation $F_{3}(x)=0$ for $x>0$.
Because we find $F_{3}(1)=-6$, $C=1$ is not a solution for arbitrary $\alpha_{3}$ and $\alpha_{4}$.
Hence, by solving $F_{3}(x)=0$, 
we may obtain non-trivial solutions with $C\neq1$ 
thanks to $M_{\mathrm{ratio}}=1$.  

When $\alpha_{3}=\alpha_{4}$, 
the equation $F_{3}(x)=0$ reduces to the quadratic equation and it has the following form:
\begin{align}
F_{2}(x) \equiv& -(4\alpha_{4} - 3) x^{2} + 2(\alpha_{4} - 3)x \nonumber \\
&+ (2\alpha_{4} - 3) = 0. \label{quadratic}
\end{align}
Furthermore, when $\alpha_{4} = \frac{3}{4}$, 
the equation $F_{2}(x)=0$ reduces to the linear equation, which is given by,
\begin{align}
F_{1}(x)\equiv - \frac{9}{2}x - \frac{3}{2}=0 . \label{linear}
\end{align}
When we solve the equation $F_{1}(x)=0$, we obtain 
\begin{align}
x=-\frac{1}{3} .
\end{align}
This is improper solution because $x$ should be positive, therefore we find
\begin{align}
\mbox{When} \ \alpha_{3}=\alpha_{4}=\frac{3}{4}, \ \mbox{no solution}. \label{no solution0}
\end{align}

\begin{figure}[htbp]
\begin{center}
\includegraphics[width=0.333\textwidth]{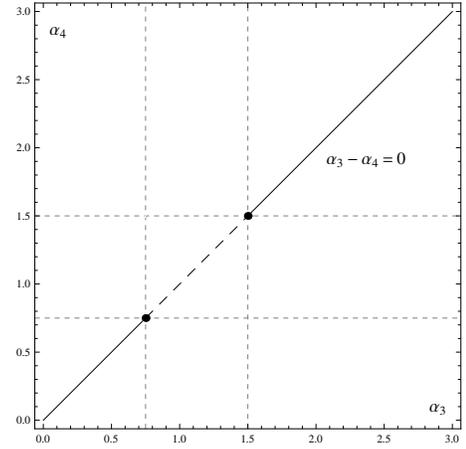}
\end{center}
\caption{
The results in the cases $F_{1}(x)=0$ and $F_{2}(x)=0$.
On the black line, we have a non-trivial solution.
However, on dashed line and its two endpoints corresponding to (\ref{no solution0}) and (\ref{no solution1}), 
there is no solution.
}
\label{fig0} 
\end{figure}

Next, we solve the equation $F_{2}(x)=0$ except for the case (\ref{no solution0}).
In order to find the number of solution, we need the discriminant,
which is given by
\begin{align}
\frac{D_{2}}{4}
=& 9 \left( \alpha_{4}-\frac{4}{3} \right)^{2} + 2 ,  \label{D2} 
\end{align}
and we have two solutions for any $x$ because $D_{2}>0$. 
However, we additionally require the condition $x>0$ on the solution, 
we find
\begin{align}
&\mbox{When} \ \alpha_{4}<\frac{3}{4} \ \mbox{or} \ \frac{3}{2} <\alpha_{4} , \ \mbox{one solution}. \\
&\mbox{When} \ \frac{3}{4}<\alpha_{4} \leqq \frac{3}{2} , \ \mbox{no solution}. \label{no solution1}
\end{align}
Getting obtained results together so far, we obtain the figure in Fig.\ref{fig0}.

\begin{figure}[htbp]
\begin{center}
\includegraphics[width=0.333\textwidth]{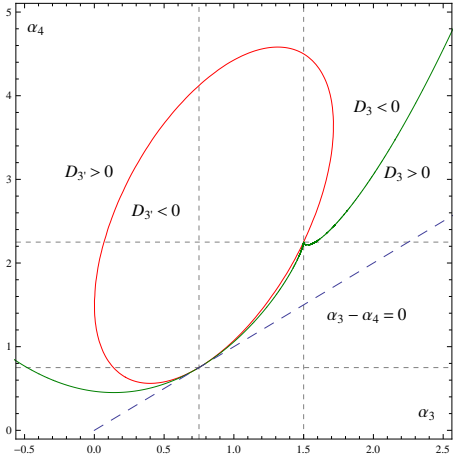}
\end{center}
\caption{l
Red line expresses $D_{3'}=0$ and green line expresses $D_{3}=0$. 
The region $D_{3}\geqq0$ is included in the region $D_{3'}\geqq0$.
Blue dashed line corresponds to $\alpha_{3}-\alpha_{4}=0$.
}
\label{fig1} 
\end{figure}

Finally, we solve the equation $F_{3}(x)=0$ when $\alpha_{3}\neq\alpha_{4}$.
The discriminant of $F_{3}(x)=0$ is given by
\begin{align}
\frac{D_{3}}{12}
=&84\alpha^{4}_{3} -48\alpha^{3}_{3}\alpha_{4} -160\alpha^{3}_{3} 
+12\alpha^{2}_{3}\alpha^{2}_{4} -84\alpha^{2}_{3}\alpha_{4} \nonumber \\
&+180\alpha^{2}_{3} +60 \alpha_{3}\alpha^{2}_{4} +144\alpha_{3}\alpha_{4} 
-108\alpha_{3} \nonumber \\
&-16\alpha^{3}_{3} -9\alpha^{2}_{4} -108\alpha_{4} +54 .
\end{align}
We also need the discriminant of $F'_{3}(x)=0$ to determine the positions of extremal value, 
where $F'_{3}(x)$ is the derivative of (\ref{cubic}).
The derivative is given by
\begin{align}
F'_{3}(x)
=& 3(\alpha_{3}-\alpha_{4})x^{2} - 2(5\alpha_{3} - \alpha_{4} - 3)x \nonumber \\
& + (\alpha_{3} + \alpha_{4} - 6) ,
\end{align}
and its discriminant is
\begin{align}
\frac{D_{3'}}{4}
=& 22\alpha^{2}_{3} -10\alpha_{3}\alpha_{4} -12\alpha{3} +4\alpha^{2}_{4} -12\alpha_{4} +9 .
\end{align}
Here, we find $D_{3'} \geqq 0$ is necessary so that $D_{3} \geqq 0$ (see Fig.\ref{fig1}), and for any $x$, we find
\begin{align}
&\quad \left\{
\begin{array}{l}
\mbox{When} \ D_{3}>0, \ \mbox{three solution} . \\
\mbox{When} \ D_{3}=0 \ \mbox{and} \ D_{3'}>0, \ \mbox{two solution} . \\
\mbox{When} \ D_{3}=0 \ \mbox{and} \ D_{3'}=0, \ \mbox{one solution} . \\
\mbox{When} \ D_{3}<0, \ \mbox{one solution} . 
\end{array}
\right. 
\end{align}
Here, by solving the simultaneous equations $D_{3}=0$ and $D_{3'}=0$, 
we obtain the points of contact, $(\alpha_{3},\alpha_{4})=(\frac{3}{4}, \frac{3}{4}), (\frac{3}{2}, \frac{9}{4})$,
and $(\alpha_{3},\alpha_{4})=(\frac{3}{4}, \frac{3}{4})$ is irrelevant to $\alpha_{3}\neq\alpha_{4}$.

We now consider the number of solutions of the equation $F_{3}(x)=0$ when $x>0$.
When $\alpha_{3}-\alpha_{4}>0$, which gives $D_{3}>0$, we find
\begin{align}
&\mbox{When} \ 3\alpha_{3} - \alpha_{4} - 3\leqq0, \mbox{one solution}. \\
&\mbox{When} \ \alpha_{3} + \alpha_{4} - 6<0 \nonumber \\
&\hphantom{\mbox{When} \ } \mbox{and} \ 3\alpha_{3} - \alpha_{4} - 3>0, \ \mbox{two solution}. \\
&\mbox{When} \ \alpha_{3} + \alpha_{4} - 6\geqq0, \ \mbox{two solution} .
\end{align}
\begin{figure}[htbp]
\begin{center}
\includegraphics[width=0.333\textwidth]{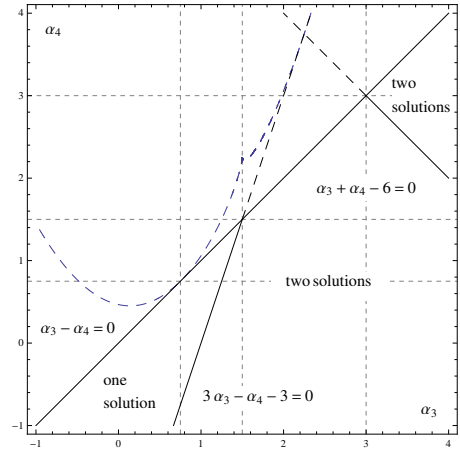}
\end{center}
\caption{
The results in case $\alpha_{3}-\alpha_{4}>0$ is drawn. 
In this region, we obtain non-trivial solutions although the number of solution is different.
}
\label{fig2-1} 
\end{figure} 

When $\alpha_{3}-\alpha_{4}<0$, we have three cases $D_{3}>0$, $D_{3}=0$ and $D_{3}<0$.
First of all, we consider the case $D_{3}>0$, and we find
\begin{align}
&\mbox{When} \ \alpha_{3}<\frac{3}{4}, \ \mbox{two solution} . \\
&\mbox{When} \ \alpha_{3}>\frac{3}{4} \nonumber \\
&\hphantom{\mbox{When} \ }  \mbox{and} \ 3\alpha_{3} - \alpha_{4} - 3\leqq0, \ \mbox{no solution} . \label{no solution2} \\
&\mbox{When} \ 3\alpha_{3} - \alpha_{4} - 3>0 \nonumber \\ 
&\hphantom{\mbox{When} \ }  \mbox{and} \ \alpha_{3} + \alpha_{4} - 6\leqq0, \ \mbox{one solution} . \\
&\mbox{When} \ 3\alpha_{3} - \alpha_{4} - 3\geqq0 \nonumber \\ 
&\hphantom{\mbox{When} \ }  \mbox{and} \ \alpha_{3} + \alpha_{4} - 6>0, \ \mbox{one solution} . \\
&\mbox{When} \ 3\alpha_{3} - \alpha_{4} - 3<0 \nonumber \\ 
&\hphantom{\mbox{When} \ }  \mbox{and} \ \alpha_{3} + \alpha_{4} - 6>0 , \ \mbox{two solution} .
\end{align}
\begin{figure}[htbp]
\begin{center}
\includegraphics[width=0.333\textwidth]{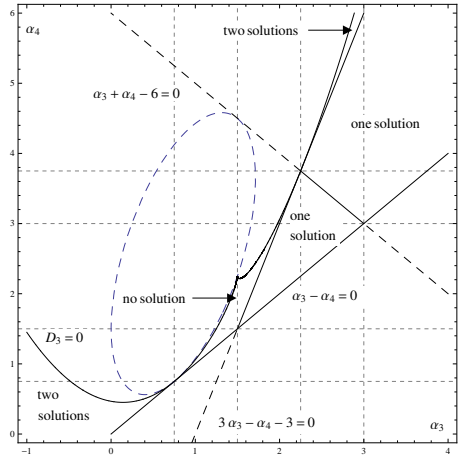}
\end{center}
\caption{
The results in case $\alpha_{3}-\alpha_{4}<0$ and $D_{3}>0$ is drawn.
In this case, we obtain a region (\ref{no solution2}) in which we can not obtain non-trivial solutions.
Except for this region, there are non-trivial solutions.
}
\label{fig2-2} 
\end{figure}

Next, we consider the case $D_{3}=0$.
In the case $D_{3'}>0$, we find
\begin{align}
&\mbox{When} \ \frac{3}{4}< \alpha_{3}<\frac{3}{2}\ \mbox{or} \ \frac{3}{2}<\alpha_{3}\leqq\frac{9}{4} , \ \mbox{no solution}. \\
&\mbox{When} \ \alpha_{3}< \frac{3}{4} \ \mbox{or} \ \frac{9}{4}< \alpha_{3}, \ \mbox{one solution}.
\end{align}
In the case $D_{3'}=0$, we find
\begin{align}
\mbox{When} \ (\alpha_{3}, \alpha_{4}) = \left( \frac{3}{2} , \frac{9}{4} \right) , \ \mbox{no solution}.
\end{align}
Getting these two results together, we find
\begin{align}
&\mbox{When} \ \frac{3}{4}< \alpha_{3}\leqq\frac{9}{4} , \ \mbox{no solution}. \label{no solution3} \\
&\mbox{When} \ \alpha_{3}< \frac{3}{4} \ \mbox{or} \ \frac{9}{4}< \alpha_{3} , \ \mbox{one solution} .
\end{align}
\begin{figure}[htbp]
\begin{center}
\includegraphics[width=0.333\textwidth]{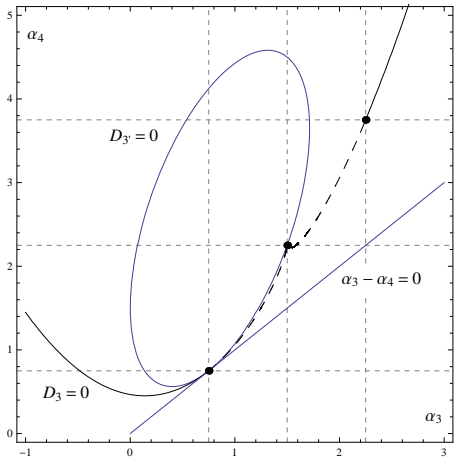}
\end{center}
\caption{
The results in case $D_{3}=0$ is drawn.
On the black line, we have a non-trivial solution.
On dashed line and points $(\alpha_{3},\alpha_{4})=(\frac{3}{2}, \frac{9}{4}), (\frac{9}{4}, \frac{15}{4})$ corresponding to (\ref{no solution3}), 
there is no non-trivial solution.
}
\label{fig2-3} 
\end{figure}

Finally, we consider the case $D_{3}<0$, 
and we find 
\begin{align}
\mbox{When} \ D_{3}<0, \ \mbox{no solution}.
\end{align}
Note that in this subsection, we omit some details of classification.
Details are given in Appendix \ref{detail of classification}.

\begin{figure}[htbp]
\begin{center}
\includegraphics[width=0.333\textwidth]{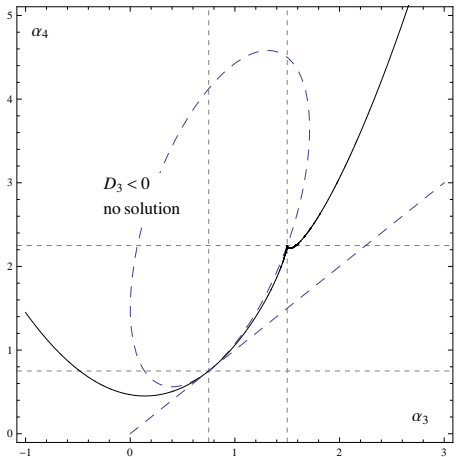}
\end{center}
\caption{
The results in case $D_{3}<0$ is drawn.
In this region, we does not have non-trivial solutions.
}
\label{fig2-4} 
\end{figure}

\subsection{Cosmological Constants}
In previous subsection, we have classified the parameter regions, 
and we find that we could obtain non-trivial solutions with $C\neq1$.
In this subsection, we discuss the properties of cosmological constants.

Substituting $M_{\mathrm{eff}}=1$ to (\ref{lambdag}) and (\ref{lambdaf}), we obtain
\begin{align}
\Lambda_{g}(C) 
=& \frac{1}{2}m^{2}_{0}\, (C-1) \nonumber \\
&\times \left [ (\alpha_{3}-\alpha_{4})C^2 + (-5\alpha_{3} + 2\alpha_{4} +3)C \right. \nonumber \\
& \left. \quad + (4\alpha_{3} - \alpha_{4} -6) \right ] , \label{lambdag'} \\
\Lambda_{f}(C)
=& \frac{1}{2}m^{2}_{0}\, \frac{C-1}{C^{3}} \nonumber \\
&\times  \left [ \alpha_{4}C^2 + (3\alpha_{3} -2\alpha_{4})C - (3\alpha_{3} - \alpha_{4} - 3) \right ] . \label{lambdaf'}
\end{align}
The magnitude of cosmological constants is proportional to square of  the mass, $m^{2}_{0}$,
and the sign depends on $\alpha_{3}$, $\alpha_{4}$ and corresponding $C$ 
which is the solution for $F_{3}(C)=0$.
Also, the two cosmological constants are related with each other by the equation $\Lambda_{g}=C^{2}\Lambda_{f}$, 
and they have same sign.

Here, we have shown that the cosmological constants vanish when $C=1$, however,
they could also vanish even if $C\neq1$. 
For convenience, we define
\begin{align}
\lambda_{g}(x)
=&(\alpha_{3}-\alpha_{4})x^2 + (-5\alpha_{3} + 2\alpha_{4} +3)x \nonumber \\
&+ (4\alpha_{3} - \alpha_{4} -6) \\
\lambda_{f}(x)
=& \alpha_{4}x^2 + (3\alpha_{3} -2\alpha_{4})x - (3\alpha_{3} - \alpha_{4} - 3) ,
\end{align}
and we find that $\lambda_{g}(1)=-3$ and $\lambda_{f}(1)=3$.
Since the case $\Lambda_{g}=\Lambda_{f}=0$ is included in $\Lambda_{g}=C^{2}\Lambda_{f}$, 
the conditions $\lambda_{g}(x)=\lambda_{f}(x)=0$ are realized in case $F_{3}(x)=0$.
Therefore, the cosmological constants vanish for some parameters 
which satisfy the conditions $\lambda_{g}(x)=\lambda_{f}(x)=0$

\subsection{In the case $M_{f}\neq M_{g}$ }
In the previous subsections,
we have considered the case where $f_{\mu \nu}=C^{2}g_{\mu \nu}$ and $M_{f}=M_{g}$,
and investigated the existence of solutions corresponding to $C$.
In this subsection, we consider the case that $M_{f}\neq M_{g}$,
therefore, the classification of solutions depends on $M^{2}_{\mathrm{ratio}}$ and becomes more complicated. 

We define a function $\bar{F}_{3}(x)$ as follows:
\begin{align}
\bar{F}_{3}(x) 
=& M^{2}_{\mathrm{ratio}}(\alpha_{3}-\alpha_{4})x^{3} \nonumber \\
& + \{ -5M^{2}_{\mathrm{ratio}}\alpha_{3} + (2M^{2}_{\mathrm{ratio}} - 1)\alpha_{4} 
+ 3M^{2}_{\mathrm{ratio}} \} x^{2} \nonumber \\
& + \{ (4M^{2}_{\mathrm{ratio}} -3)\alpha_{3} - (M^{2}_{\mathrm{ratio}} - 2)\alpha_{4} \nonumber \\
& - 6M^{2}_{\mathrm{ratio}} \}x + (3\alpha_{3} - \alpha_{4} - 3) .
\end{align}
When $\alpha_{3}=\alpha_{4}$, 
the equation $\bar{F}_{3}(x)=0$ reduces to the quadratic equation as well as the case $M_{f}=M_{g}$, 
which is given by
\begin{align}
\bar{F}_{2}(x)=& \{ - (3M^{2}_{\mathrm{ratio}} + 1)\alpha_{4} + 3M^{2}_{\mathrm{ratio}} \} x^{2} \nonumber \\
& + \{ (3M^{2}_{\mathrm{ratio}} - 1)\alpha_{4} - 6M^{2}_{\mathrm{ratio}} \}x \nonumber \\
& + (2\alpha_{4} - 3) = 0 .
\end{align}
Furthermore, when $\alpha_{4} = 3M^{2}_{\mathrm{ratio}}/(3M^{2}_{\mathrm{ratio}} + 1)$, 
the equation $\bar{F}_{2}(x)=0$ reduces to the linear equation and it is given by
\begin{align}
\bar{F}_{1}(x)=& - \frac{9M^{2}_{\mathrm{ratio}}(M^{2}_{\mathrm{ratio}}+1)}{3M^{2}_{\mathrm{ratio}}+1}x 
- \frac{3(M^{2}_{\mathrm{ratio}}+1)}{3M^{2}_{\mathrm{ratio}}+1} \nonumber \\
=& 0.
\end{align}
Solving the equation $\bar{F}_{1}(x)=0$, we obtain
\begin{align}
x=-\frac{1}{3M^{2}_{\mathrm{ratio}}} .
\end{align}
This is improper solution, and we does not obtain non-trivial solutions.
In the following, we only consider the equation $F_{2}(x)=0$ because it is tedious to investigate all of the cases.

For the equation $\bar{F}_{2}(x)=0$, its discriminant is given by
\begin{align}
\bar{D}_{2}
=& 3(M^{2}_{\mathrm{ratio}} + 1) \nonumber \\
&\times [ 3(M^{2}_{\mathrm{ratio}} + 1)\alpha_{4}^{2}  \nonumber \\
& \quad - 4(3M^{2}_{\mathrm{ratio}} + 1)\alpha_{4}+ 12M^{2}_{\mathrm{ratio}} ] . \label{D2-2} 
\end{align}
Since the discriminant apparently depends on $M^{2}_{\mathrm{ratio}}$, 
the existence of solutions also depends on  $M^{2}_{\mathrm{ratio}}$.
In order to analyze the sign of (\ref{D2-2}), 
we define a function $f_{2}(\alpha_{4})$ as follows:
\begin{align}
f_{2}(\alpha_{4})
=& 3(M^{2}_{\mathrm{ratio}} + 1)\alpha_{4}^{2} \nonumber \\
& - 4(3M^{2}_{\mathrm{ratio}} + 1)\alpha_{4} + 12M^{2}_{\mathrm{ratio}} .
\end{align}
For the equation $f_{2}(\alpha_{4})=0$, the discriminant is given by
\begin{align}
\frac{\bar{D}_{2'}}{4} 
=& 4(3M^{2}_{\mathrm{ratio}} + 1)^{2} - 36(M^{2}_{\mathrm{ratio}} + 1) M^{2}_{\mathrm{ratio}} \nonumber \\
=& -12 \left( M^{2}_{\mathrm{ratio}} - \frac{1}{3} \right) \label{D2alpha4} \, .
\end{align}
This shows that when $ M^{2}_{\mathrm{ratio}} < \frac{1}{3}$, 
$\bar{D}_{2}$ can be negative according to the value of $\alpha_{4}$,
which tells that there is no solution for $\bar{F}_{2}(x)=0$.

\section{Physical Interpretation of $f_{\mu \nu}$ and the Planck mass scales}

Although we have solved the equations of motion and investigated the properties of solutions,
we have not considered the physical meaning of “two” metric.
While the bigravity theory contains two metric tensor fields as expressed up until now, 
it could be natural to consider that physical metric which defines the length or area should be only one of the two tensor fields.
In that sense, bigravity is the theory which describes coupling between gravity and a symmetric rank-two tensor field
\cite{Akrami:2013ffa}, that is,
$g_{\mu \nu}$ could be the physical metric which describes gravity and $f_{\mu \nu}$ could be just a tensor field.

In this section, however,
we may assume that both metrics have actually geometrical meaning as a measure 
and we consider how we should interpret the two metrics.
Especially, we consider the two Planck mass scales $M_{f}$ and $M_{g}$ 
by combining the above assumption and the condition $f_{\mu \nu}=C^{2}g_{\mu \nu}$.

As is well known, the Planck scale $M_{\mathrm{Planck}}$ is defined as a scale 
where the Compton wavelength is comparable with the Schwarzschild radius in the general relativity, 
that is,
\begin{align}
\frac{\hbar}{M_{\mathrm{Planck}}c} \sim \frac{2GM_{\mathrm{Planck}}}{c^{2}} , \label{planck mass scales}
\end{align}
where $c$ is the speed of light.
Therefore, if the metric $f_{\mu \nu}$ has also geometrical meaning, 
the Planck scale for $f$ can be also evaluated by above definition.
The two scales $M_{g}$ and $M_{f}$ are defined as follows:
\begin{align}
M_{g} \sim \sqrt{\frac{\hbar c}{G_{g}}} , \quad
M_{f} \sim \sqrt{\frac{\hbar c}{G_{f}}} ,
\end{align}
where, $G_{g}$ and $G_{f}$ are the gravitational constants for $g$ and $f$, respectively,
In order to find the relation between $M_{g}$ and $M_{f}$, 
we need to know how the two physical constants $G_{g}$ and $G_{f}$ are related to each other 
by putting $f_{\mu \nu}=C^{2}g_{\mu \nu}$.

We now consider the Schwarzschild solution, then the metric $f_{\mu \nu}$ is expressed as follows:
\begin{align}
&f_{\mu \nu} {\rm d}x^{\mu} {\rm d}x^{\nu} \nonumber \\
&= C^{2}g_{\mu \nu} {\rm d}x^{\mu} {\rm d}x^{\nu} \nonumber \\
&= -C^{2} \left( 1-\frac{2m_{g}G_{g}}{c^{2}r} \right) {\rm d}t^{2} 
+ C^{2} \left( 1-\frac{2m_{g}G_{g}}{c^{2}r} \right)^{-1} {\rm d}r^{2} \nonumber \\
&\qquad + C^{2}r^{2} {\rm d}\Omega^{2} .
\end{align}
Here, $m_{g}$ is a mass which are measured by a observer in the space-time described by $g_{\mu \nu}$.
Because we need the asymptotically flat form of $f_{\mu \nu}$,
we consider the coordinate transformation, $(t',r')=(Ct,Cr)$.
In the coordinates $(t',r')$, $f_{\mu \nu}$ is given by 
\begin{align}
&f_{\mu \nu} {\rm d}x'^{\mu} {\rm d}x'^{\nu} \nonumber \\
&\quad \equiv - \left( 1-\frac{2m_{f}G_{f}}{r'} \right) {\rm d}t'^{2}  
+  \left( 1-\frac{2M_{f}m_{f}}{r'} \right)^{-1} {\rm d}r'^{2} \nonumber \\
&\quad \quad + r'^{2} {\rm d}\Omega^{2} .
\end{align}
Here, we define
\begin{align}
Cm_{g}G_{g}=m_{f}G_{f}, \label{gravitational potential}
\end{align}
where $m_{f}$ is a mass which are measured by a observer in the space-time described by $f_{\mu \nu}$.

On the other hand, 
the Compton wavelength measured in the space-time described by $g_{\mu \nu}$ is also different from that by $f_{\mu \nu}$
because of the coordinate transformation.
The relation between the two Compton wavelength is given by
\begin{align}
\frac{C\hbar}{m_{g}c}=\frac{\hbar}{m_{f}c} . \label{wavelengths}
\end{align}
Combining (\ref{gravitational potential}) and (\ref{wavelengths}), 
we obtain the relation between two Planck mass scales as follows:
\begin{align}
M^{2}_{\mathrm{ratio}}=\left( \frac{M_{f}}{M_{g}} \right)^{2} \sim \frac{1}{C^{2}} .
\end{align}
This is the case $M_{f}\neq M_{g}$, furthermore, 
$M_{f}$ and $M_{g}$ are related through $C$ rather than the independent theoretical parameters.

\section{Black Hole Entropy for Spherically-symmetric solution}

\subsection{Black Hole Entropy from the Noether Current}
In this section, we consider the black hole entropy in bigravity.
As shown in previous section, we find that the dynamics of bigravity is same as the one of the general relativity 
by assuming the condition $f=C^{2}g$.
Therefore, we can utilize the black hole solution of the general relativity to evaluate the entropy.
In order to evaluate the entropy in bigravity, we use the recently proposed method by Majhi and Padmanabhan 
\cite{Katsuragawa:2013bma,Majhi:2011ws,Majhi:2012tf,Majhi:2012nq}.
In this subsection, we summarize this procedure.

Let us consider a general surface term as follows: 
\begin{align}
I_{B} 
=& \frac{1}{16 \pi G} \int_{\mathcal{\partial M}} \, d^{n-1}x \sqrt{\sigma} \mathcal{L}_{B} \nonumber \\
=& \frac{1}{16 \pi G} \int_{\mathcal{M}} \, d^{n}x \sqrt{g}\nabla_{a}(\mathcal{L}_{B} N^{a}).
\end{align}
Here, $N^{a}$ is a unit normal vector of the boundary $\partial \mathcal{M}$, 
$g_{\mu \nu}$ is the bulk metric, and $\sigma_{\mu \nu}$ is the induced boundary metric.
For the Lagrangian density $ \sqrt{g} \mathcal{L} = \sqrt{g}\nabla_{a}(\mathcal{L}_{B} N^{a})$, 
the conserved Noether current corresponding to differmorphism $ x^{a} \rightarrow x^{a} + \xi^{a} $ 
is given by (see the appendix in \cite{Majhi:2012tf})
\begin{align}
J^{a}[\xi] 
= \nabla_{b}J^{ab}[\xi] 
= \frac{1}{16 \pi G} \nabla_{b}
\left[ \mathcal{L}_{B} \left( \xi^{a}N^{b} - \xi^{b}N^{a} \right) \right].
\end{align}
Here, $J^{ab}$ is the Noether potential, 
and the corresponding charge is defined as
\begin{align}
Q[\xi] = \frac{1}{2} \int_{\partial \Sigma} \, \sqrt{h} d\Sigma_{ab}J^{ab}.
\end{align}
Here, $ d\Sigma_{ab} = - d^{n-2}x \left( N_{a}M_{b} - N_{b}M_{a} \right)$ is 
the surface element of the $(n-2)$-dimensional 
surface $\partial \Sigma$, and $h_{ab}$ is the corresponding induced metric.
We now choose the unit normal vectors $N^{a}$ and $M^{b}$ as spacelike and timelike, respectively.
In the following disucussion, we assume $\Sigma$ exists near the horizon of a black hole. 

Next, we define the Lie bracket for the charges as follows: 
\begin{align}
[Q_{1}, Q_{2}] 
=& \left( \delta_{\xi_{1}} Q[\xi_{2}] - \delta_{\xi_{2}} Q[\xi_{1}] \right) \nonumber \\
=& \int_{\partial \Sigma} \, \sqrt{h} d\Sigma_{ab} 
\left( \xi^{a}_{2} J^{b}[\xi_{1}] - \xi^{a}_{1} J^{b}[\xi_{2}] \right), 
\label{lie bracket}
\end{align}
which leads to the Virasoro algebra with central extension as we will see later. 
By using the deduced central charge and the Cardy formula, one can find black hole entropy.

To derive the Noether charge and the Virasoro algebra, we need to identify appropriate diffeormorphisms, that is, 
the related vector field $\xi^{a}$.
In this work, we consider static-spherical black holes, where the metric has the following form: 
\begin{align}
ds^{2} = -f(r)dt^{2} + \frac{1}{f(r)}dr^{2} + r^{2}\Omega_{ij}(x)dx^{i}dx^{j}.
\label{BHmetric}
\end{align}
Here, $\Omega_{ij}(x)$ is the $(n-2)$-dimensional space, 
and $h_{ij}=r^{2}\Omega_{ij}(x)$.
The horizon exist at $r=r_{h}$, where $f(r_{h})=0$.
For the metric (\ref{BHmetric}), the two normal vectors $N^{a}$ and $M^{a}$ are given by 
\begin{align}
&N^{a} = \left( 0, \sqrt{f(\rho + r_{h})}, 0, \cdots , 0 \right) \nonumber \\
&M^{a} = \left( \frac{1}{\sqrt{f(\rho + r_{h})}}, 0, \cdots , 0 \right) .
\end{align}
Here, $\rho$ is defined by $r = \rho + h_{h}$ for convenience, and in the near horizon limit, 
we find $\rho \rightarrow 0$.
Then, the metric has the following form:
\begin{align}
ds^{2} =& -f(\rho + r_{h})dt^{2} + \frac{1}{f(\rho + r_{h})}d\rho^{2} \nonumber \\
& \hphantom{-f(\rho + r_{h})dt^{2}} + (\rho + r_{h})^{2}\Omega_{ij}(x)dx^{i}dx^{j} .
\end{align}
Furthermore, we introduce the Bondi-like coordinates, 
\begin{align}
du = dt - \frac{d\rho}{f(\rho + r_{h})}, \label{bondi-like coordinate}
\end{align} 
and the metric is transformed as 
\begin{align}
ds^{2} =& - f(\rho + r_{h})du^{2} - 2dud\rho \nonumber \\
& \hphantom{- f(\rho + r_{h})du^{2}} + (\rho + r_{h})^{2}\Omega_{ij}(x)dx^{i}dx^{j} .
\end{align}
We choose the vector fields $\xi^{a}$ so that the vector fields keep 
the horizon structure invariant.
Then, we now solve the Killing equations for above metric:
\begin{align}
&\mathcal{L}_{\xi} g_{\rho \rho} = -2\partial_{\rho} \xi^{u} = 0 ,\nonumber \\
&\mathcal{L}_{\xi} g_{u \rho}
= - \partial_{u} \xi^{u} - f(\rho + r_{h}) \partial_{\rho} \xi^{u} - \partial_{\rho} \xi^{\rho} = 0.
\end{align}
The solutions of the above equations are given by 
\begin{align}
\xi^{u} = F(u,x) ,\quad 
\xi^{\rho} = - \rho \partial_{u}F(u,x) .
\end{align}
The remaining condition $\mathcal{L}_{\xi}g_{uu}=0$ is automatically satisfied near the horizon because the 
above solutions lead to $\mathcal{L}_{\xi}g_{uu} = \mathcal{O}(\rho)$ and $\rho \rightarrow 0$ at the horizon. 
Expressing the results in the original coordinates $(t, \rho)$, we obtain
\begin{align}
&\xi^{t} = T - \frac{\rho}{f(\rho + r_{h})} \partial_{t}T ,\quad 
\xi^{\rho} = - \rho \partial_{t}T , \nonumber \\
&T(t, \rho, x) = F(u, x) .
\end{align}
Since we have the appropriate vector fields $\xi^{a}$ for a given $T$, we can calculate the charge $Q$.

Finally, expanding $T$ in terms of a set of basis functions $T_{m}$ with
\begin{align}
T = \sum_{m} A_{m}T_{m} ,\quad A^{*}_{m} = A_{-m},
\end{align}
we obtain corresponding expansions for $\xi^{a}$ and $Q$ in terms of $\xi^{a}_{m}$ and $Q_{m}$ defined by $T_{m}$.
We choose $T_{m}$ to be the basis so that the resulting $\xi^{a}_{m}$ obeys the algebra isomorphic 
to Diff $S^{1}$, 
\begin{align}
i\{ \xi_{m}, \xi_{n} \}^{a} = (m-n) \xi^{a}_{m+n},
\end{align}
with $\{ , \}$ as the Lie bracket.
Such a $T_{m}$ can be achieved by the choice
\begin{align}
T_{m} = \frac{1}{\alpha}\mathrm{exp}[im(\alpha t + g(\rho) + p \cdot x)] \label{T_{m}}.
\end{align}
Here, $\alpha$ is a constant, $p$ is an integer, and $g(\rho)$ is a function that is regular at the horizon.
Note that $\alpha$ is arbitrary in this approach, which will not affect the final results.

\subsection{Entropy for Spherically Symmetric Black hole in Bigravity }

Using the previous procedure and the black hole solution, we evaluate the black hole entropy.
At first, we need to calculate surface term of the bigravity action $\mathcal{L}_{B}$ 
and the vector field $\xi^{a}$ related to the diffeormophism, which leaves the horizon structure invariant.

Since the interaction term does not include any derivative terms, the contribution to the surface term does not appear.
Therefore, the surface term is generally obtained by two Gibbons-Hawking terms from Ricci scalar $R(g)$ and $R(f)$,
\begin{align}
\sqrt{\sigma}\mathcal{L}_{B} = 2\sqrt{\sigma_{g}}K(g) + 2\sqrt{\sigma_{f}}K(f) , \label{boundary_bigravity}
\end{align}
with $K = - \nabla_{a}N^{a}$ as the trace of the extrinsic curvature of the boundary,
and indices for $g$ and $f$, respectively.

When we consider the Schwarzschild solutions, $f_{\mu \nu} = C^{2}g_{\mu \nu}$ and $f(r) = 1 - \frac{2M}{r}$ 
with the horizon at $r_{h} = 2M$.
Metric tensor $g_{\mu \nu}$ corresponding to the coordinates $(t, \rho)$ is given by 
\begin{align}
ds^{2} =& - \frac{\rho}{\rho + 2M} dt^{2} + \frac{\rho + 2M}{\rho}d\rho^{2} \nonumber \\
& \qquad \qquad + (\rho + 2M)^{2}(d\theta^{2} + \sin^{2} \theta d\phi^{2}) .
\end{align}
The Bondi-like coordinate transformation (\ref{bondi-like coordinate}) is defined as
\begin{align}
du = dt - \frac{2M + \rho}{\rho} .
\end{align}
In this coordinate system, the metric is expressed as 
\begin{align}
ds^{2} =& - \frac{\rho}{\rho + 2M} du^{2} - 2dud\rho \nonumber \\
& \qquad \qquad + (\rho + 2M)^{2}(d\theta^{2} + \sin^{2} \theta d\phi^{2}).
\end{align}
The vector fields $\xi^{a}$ in the original coordinates $(t, \rho)$ have the following expressions: 
\begin{align}
\xi^{t} = T - (\rho + 2M) \partial_{t}T ,\quad \xi^{\rho} = - \rho \partial_{t}T . \label{diff_vector}
\end{align}
Note that above diffeomorphism (\ref{diff_vector}) also keeps the horizon strucuture invariant for $f_{\mu \nu}$
because $f_{\mu \nu}=C^{2}g_{\mu \nu}$, while we have one diffeomorphism for two metrics in bigravity.

We now calculate the Noether current and the Virasoro algebra.
The spacelike normal vectors for $g$ and $f$ at the horizon are
\begin{align}
&N^{a}(g) = \left( 0, \sqrt{ \frac{\rho}{\rho + 2M}}, 0, 0 \right) \nonumber \\
&N^{a}(f) = \left( 0, \frac{1}{C}\sqrt{ \frac{\rho}{\rho + 2M}}, 0, 0 \right) ,
\end{align}
and we find 
\begin{align}
K(f)= \frac{1}{C}K(g) , 
\end{align}
because covariant derivatives are invariant by putting $f=C^{2}g$.
Also, $\sigma$ is three-dimensional induced metric and we find
\begin{align}
\sqrt{\sigma_{f}}=C^{3}\sqrt{\sigma_{g}} ,
\end{align}
as a result, the surface term (\ref{boundary_bigravity}) is
\begin{align}
\sqrt{\sigma}\mathcal{L}_{B} = 2(1+C^{2})\sqrt{\sigma_{g}}K(g) .
\end{align}
For Schwarzschild solution, the Gibbons-Hawking term is given by 
\begin{align}
K(g) = -\frac{2 \rho + M}{\sqrt{\rho} \, (\rho + 2M)^{3/2}}.
\end{align}
The charge $Q$ in the near-horizon limit $\rho \rightarrow 0$ is given by
\begin{align}
Q[\xi] = \frac{(1+C^{2})}{8 \pi G} \int_{\mathcal{H}} \, \sqrt{h}d^{2}x 
[\kappa T - \frac{1}{2} \partial_{t}T] ,
\label{charge}
\end{align}
where $\kappa$ is the surface gravity of the black hole, $\kappa = \frac{1}{4M}$.

Finally, we calculate the central charge with the appropriate expansion of $T$.
For $T = T_{m}$, $T_{n}$, the Lie bracket of the charges $Q_{m}$ and $Q_{n}$ (\ref{lie bracket}) 
is given by 
\begin{align}
[Q_{m}, Q_{n}] 
=& \frac{(1+C^{2})}{8 \pi GM}\int_{\mathcal{H}} \, d^{2}x
\left[ \kappa ( T_{m} \partial_{t}T_{n} - T_{n} \partial_{t}T_{m} ) \right. \nonumber \\
& - \frac{1}{2}( T_{m} \partial^{2}_{t}T_{n} - T_{n} \partial^{2}_{t}T_{m} ) \nonumber \\
& + \frac{1}{4 \kappa} \left.  ( \partial_{t} T_{m} \partial^{2}_{t}T_{n} - \partial_{t} T_{n} \partial^{2}_{t}T_{m}) \right] .
\label{lie bracket for charge}
\end{align}
We now substitute $T_{m}$ chosen in the previous subsection (\ref{T_{m}}) into Eqs.(\ref{charge}) and 
(\ref{lie bracket for charge}) 
and implement the integration over a cross-section area $A$, 
and we obtain 
\begin{align}
Q_{m} =&\frac{(1+C^{2})A}{8 \pi G} \frac{\kappa}{\alpha} \delta_{m,0} ,\\
[Q_{m}, Q_{n}] =& - \frac{i (1+C^{2})A}{8 \pi G} \frac{\kappa}{\alpha} (m-n) \delta_{m+n,0} \nonumber \\
& - im^{3}\frac{(1+C^{2})A}{16 \pi G} \frac{\alpha}{\kappa} \delta_{m+n,0} .
\end{align}
Therefore, we find that the central term in the algebra is given by 
\begin{align}
K[\xi_{m}, \xi_{n}] =& [Q_{m}, Q_{n}] + i(m-n)Q_{m+n} \nonumber \\
=& -im^{3} \frac{(1+C^{2})A}{16 \pi G}\frac{\alpha}{\kappa} \delta_{m+n,0} .
\end{align}
From the central term, we can read off the central charge $C_{c}$ 
and the zero mode energy $Q_{0}$ as follows:
\begin{align}
\frac{C_{c}}{12} = \frac{(1+C^{2})A}{16 \pi G}\frac{\alpha}{\kappa} , \quad 
Q_{0} = \frac{(1+C^{2})A}{8 \pi G} \frac{\kappa}{\alpha}  .
\end{align}
Using the Cardy formula
\cite{Cardy:1986ie,Bloete:1986qm,Carlip:1998qw}, 
we eventually obtain the entropy
\begin{align}
S = 2 \pi \sqrt{\frac{C_{c}Q_{0}}{6}} = (1+C^{2}) \, \frac{A}{4G} .
\end{align}
This is different from the Bekenstein-Hawking entropy in the Einstein gravity by the factor $1+C^{2}$,
and this result is consistent with the result of BTZ black hole in three-dimensional bigravity \cite{Banados:2011np}.

\section{Summary and Discussion}
We have studied the properties of general model in bigravity with condition $f_{\mu \nu}=C^{2}g_{\mu \nu}$.
In this condition, 
we have shown that the structure of dynamics is not changed that of the general relativity,
the solutions of bigravity are also that of the general relativity as well.
We have also investigated the parameter region with condition $M_{f}=M_{g}$, 
which we can obtain the non-trivial solution $C\neq1$.

Then, we have shown that the entropy in bigravity is obtained by the sum of two entropies from two Ricci scalar,
and the effect of massive spin-$2$ field is identical form with the Bekenstien-Hawking entropy.
In evaluating the black hole entropy, 
we could ignore the interaction term 
because the approach in this work utilize the surface term of the action.
The remaining terms corresponding to the Noether current are simply the two traces of the extrinsic curvature, 
it makes simple and easy to calculate the entropy. 
Thus it is the advantage of this approach and we can find that the entropy is independent of a difference in the models.

Furthermore, we find that the entropy is given by the sum of two entropies for general black hole solution, 
that is, the case where two metric are independent;
because total Noether current is given by the sum of two Noether current for $g$ and $f$,
total entropy is also given by the sum after the integration over the area of horizon.

Regarding the parameter region, 
we can find that roughly, there is only trivial solution in the region where $\alpha_{4}$ is large and $|\alpha_{3}|$ is small.
And there is always non-trivial solutions in the region where $\alpha_{4}$ is negative or $|\alpha_{3}|$ is large.
Here, we consider some specific parameters and its properties.
For $(\alpha_{3},\alpha_{4})=(1,1)$, what we call the minimal model, 
we does not have non-trivial solution, and then, cosmological constants vanish.
When we consider the black hole entropy for the minimal model,
we obtain the double portion of the Bekenstein-Hawking entropy.
In order to obtain non-trivial solution, we need to consider other parameter combination.
For example, we can propose not minimal but simple combinations as follows:
\begin{align}
(\alpha_{3},\alpha_{4})=(-1,-1), \ (-1,1), \ (1,-1) ,
\end{align}
these combinations produce non-trivial solution.
\begin{figure}[htbp]
\begin{center}
\includegraphics[width=0.333\textwidth]{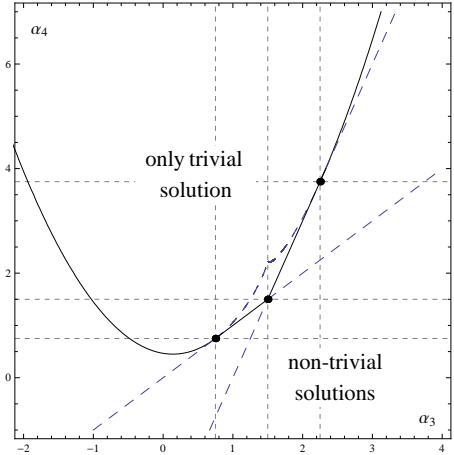}
\end{center}
\caption{
All of the results obtained in section.\ref{classification} is drawn.
In the region where $\alpha_{4}$ is large and $|\alpha_{3}|$ is small,
we does not obtain non-trivial solution.
The minimal model $(\alpha_{3},\alpha_{4})=(1,1)$ is included in the region where there is only trivial solution.
}
\label{fig3} 
\end{figure}

Note that the results are based on the condition $f_{\mu \nu}=C^{2}g_{\mu \nu}$,
we would obtain various solutions in the minimal model on different conditions.
Furthermore, we have not considered ordinary matter couplings in this work,
so we need more analysis with different condition and matter couplings.

It is interesting that we consider the relation $f_{\mu \nu}=\mathrm{exp}[2\phi] g_{\mu \nu}$ with a function $\phi(x)$
and spherical symmetry.
If we find a consistent solution, we could obtain the black hole solutions with different horizons.
However, concerning the black hole entropy, if the horizons are not located at the same position,
we cannot use the procedure in this work to evaluate the entropy.
Besides this problem, recent papers \cite{Banados:2011hk,Deffayet:2011rh} indicate 
that the horizons for black holes in bigravity must be located at the same position.

In order to find the properties without fixing the model, it is also important to consider the Bianchi identity.
The solutions must satisfy not only the equations of motion but the Bianchi identity, 
this condition constraints the form of the solution.
The interaction terms should produce suitable term to vanish when the covariant derivative operate.

\section*{Acknowledgements}

Author is partially supported by the Nagoya University Program for Leading Graduate Schools funded by the Ministry 
of Education of the Japanese Government under the program number N01.
I am deeply grateful to my supervisor, Shin'ichi Nojiri, for constructive advice and discussion.
I would like to offer my special thanks to colleagues, Yusuke Kokusho, Hiroshi Suenobu also. 
\appendix
\section{Details in Classification of Solutions}\label{detail of classification}

In this section, we consider the classification of solutions in detail.
For the equation $F_{2}(x)=0$, the classification is given by three criteria:
convex or concave, the position of axis, the value of $F_{2}(0)$.

The discriminant of $F_{2}(x)=0$ is always positive, and classification is given as follows:
\begin{align}
&\mbox{In case} \ -(4\alpha_{4} - 3)>0, \nonumber \\ 
& \quad \left\{
\begin{array}{l}
\mbox{When} \ \alpha_{4} - 3<0 \ \mbox{and} \ 2\alpha_{4} - 3>0 , \ \mbox{two solutions}. \\
\mbox{When} \ \alpha_{4} - 3 > 0\ \mbox{and} \ 2\alpha_{4} - 3 \geq 0 , \ \mbox{no solution}. \\  
\mbox{Otherwise} , \ \mbox{one solution}.
\end{array}
\right.  \\[2mm]
&\mbox{In case} \ -(4\alpha_{4} - 3)<0 , \nonumber \\ 
& \quad \left\{
\begin{array}{l}
\mbox{When} \ \alpha_{4} - 3>0 \ \mbox{and} \ 2\alpha_{4} - 3<0 , \ \mbox{two solutions}. \\ 
\mbox{When} \ \alpha_{4} - 3 < 0 \ \mbox{and} \ 2\alpha_{4} - 3 \leq 0 , \ \mbox{no solution}. \\ 
\mbox{Otherwise}  , \ \mbox{one solution}.
\end{array}
\right. 
\end{align}

Next, we consider the equation $F_{3}(x)=0$.
In order to classify the solution, we need to know the position and the number of extremal values.
So, we need the derivative of $F_{3}(x)$, 
and we obtain same criteria of $F_{2}(x)=0$ because $F'_{3}(x)$ is quadratic function.

When $\alpha_{3}-\alpha_{4}>0$, the discriminant $D_{3}$ is positive, and classification is given by
\begin{align}
&\left\{
\begin{array}{l}
\mbox{When} \ 3\alpha_{3} - \alpha_{4} - 3<0 , \ - 2(5\alpha_{3} - \alpha_{4} - 3)<0 \\ 
\quad \mbox{and} \ \alpha_{3} + \alpha_{4} - 6>0 , \ \mbox{three solution}. \\[2mm]
\mbox{When} \ 3\alpha_{3} - \alpha_{4} - 3\geqq0 , \ - 2(5\alpha_{3} - \alpha_{4} - 3)<0 \\ 
\quad \mbox{and} \ \alpha_{3} + \alpha_{4} - 6\geqq0 , \ \mbox{two solution}. \\[2mm]
\mbox{When} \ 3\alpha_{3} - \alpha_{4} - 3>0 \\ 
\quad \mbox{and} \ \alpha_{3} + \alpha_{4} - 6<0 , \ \mbox{two solution}. \\[2mm]
\mbox{When} \ 3\alpha_{3} - \alpha_{4} - 3\leqq0 \\ 
\quad \mbox{and} \ \alpha_{3} + \alpha_{4} - 6<0 , \ \mbox{one solution}. \\[2mm]
\mbox{When} \ 3\alpha_{3} - \alpha_{4} - 3<0 , \ - 2(5\alpha_{3} - \alpha_{4} - 3)>0 \\ 
\quad \mbox{and} \ \alpha_{3} + \alpha_{4} - 6\geqq0 , \ \mbox{one solution}. \\[2mm]
\mbox{When} \ 3\alpha_{3} - \alpha_{4} - 3\geqq0 , \ - 2(5\alpha_{3} - \alpha_{4} - 3)>0 \\ 
\quad \mbox{and} \ \alpha_{3} + \alpha_{4} - 6>0 , \ \mbox{no solution}.
\end{array}
\right. 
\end{align}

When $\alpha_{3}-\alpha_{4}<0$, $D_{3}$ can be positive, zero and negative, and classification is given as follows, respectively:
\begin{align}
&\mbox{In case} \ D_{3}>0, \nonumber \\
&\quad \left\{
\begin{array}{l}
\mbox{When} \ 3\alpha_{3} - \alpha_{4} - 3>0 , \ - 2(5\alpha_{3} - \alpha_{4} - 3)>0 \\ 
\quad  \mbox{and} \ \alpha_{3} + \alpha_{4} - 6<0 , \ \mbox{three solution}. \\[2mm]
\mbox{When} \ 3\alpha_{3} - \alpha_{4} - 3\leqq0 , \ - 2(5\alpha_{3} - \alpha_{4} - 3)>0 \\ 
\quad  \mbox{and} \ \alpha_{3} + \alpha_{4} - 6\leqq0 , \ \mbox{two solution}. \\[2mm]
\mbox{When} \ 3\alpha_{3} - \alpha_{4} - 3<0 \\ 
\quad \mbox{and} \ \alpha_{3} + \alpha_{4} - 6>0   , \ \mbox{two solution}. \\[2mm]
\mbox{When} \ 3\alpha_{3} - \alpha_{4} - 3\geqq0  \\ 
\quad \mbox{and} \ \alpha_{3} + \alpha_{4} - 6>0  , \ \mbox{one solution}. \\[2mm]
\mbox{When} \ 3\alpha_{3} - \alpha_{4} - 3>0 , \ - 2(5\alpha_{3} - \alpha_{4} - 3)<0 \\ 
\quad  \mbox{and} \ \alpha_{3} + \alpha_{4} - 6\leqq0 , \ \mbox{one solution}. \\[2mm]
\mbox{When} \ 3\alpha_{3} - \alpha_{4} - 3\leqq0 , \ - 2(5\alpha_{3} - \alpha_{4} - 3)<0 \\ 
\quad  \mbox{and} \ \alpha_{3} + \alpha_{4} - 6<0 , \ \mbox{no solution}.
\end{array}
\right.  
\end{align}

\begin{align}
&\mbox{In case} \ D_{3}=0, \nonumber \\
&\quad \left\{
\begin{array}{l}
\mbox{When} \ 3\alpha_{3} - \alpha_{4} - 3>0 , \ - 2(5\alpha_{3} - \alpha_{4} - 3)>0 \\ 
\quad  \mbox{and} \ \alpha_{3} + \alpha_{4} - 6<0 , \ \mbox{two solution}. \\[2mm]
\mbox{When} \ 3\alpha_{3} - \alpha_{4} - 3\leqq0 , \ - 2(5\alpha_{3} - \alpha_{4} - 3)<0 \\ 
\quad  \mbox{and} \ \alpha_{3} + \alpha_{4} - 6\leqq0 , \ \mbox{no solution}. \\[2mm]
\mbox{Otherwise} , \ \mbox{one solution}.
\end{array}
\right. 
\end{align}

\begin{align}
&\mbox{In case} \ D_{3}<0, \nonumber \\
&\quad \left\{
\begin{array}{l}
\mbox{When} \ 3\alpha_{3} - \alpha_{4} - 3>0 , \ \mbox{one solution}. \\[2mm]
\mbox{When} \ 3\alpha_{3} - \alpha_{4} - 3\leqq0 , \ \mbox{no solution} .
\end{array}
\right. 
\end{align}

\end{document}